\documentclass[10pt,a4paper]{article}

\usepackage[english]{babel}
\usepackage[cp1252]{inputenc}
\usepackage{color}
\usepackage{pstricks}
\usepackage{multicol}
\usepackage{anysize}
\usepackage{fancyhdr}
\usepackage{wasysym}
\usepackage{lettrine}
\usepackage{type1cm}
\usepackage{graphicx}
\usepackage[small,bf,centerlast]{caption}
\usepackage{bold-extra}

\usepackage{pdfpages}

\marginsize{1.5cm}{1.5cm}{0.5cm}{2cm}

\definecolor{blue4}{rgb}{0,0,0.5}
\definecolor{indianred3}{rgb}{0.8,0.33,0.33}
\definecolor{gainsboro}{rgb}{0.86,0.86,0.86}
\definecolor{dodgerblue4}{rgb}{0.06,0.31,0.55}

\makeatletter

\newenvironment{figurehere}
{\def\@captype{figure}}
{}
\makeatother

\begin{document}

\begin{center}
\textbf{\LARGE Capturing coevolutionary signals in repeat proteins}\\
\end{center}

\medskip

\begin{center}
\begin{tabular}{ccccc}
Roc\'io Espada$^\ddagger$& R. Gonzalo Parra$^\ddagger$ & Thierry Mora$^\S$ & Aleksandra M. Walczak$^\star$&Diego Ferreiro$^{\ddagger \**}$\\
\end{tabular}
\end{center}

\begin{minipage}[c]{16cm}
\footnotesize{
\noindent$^\ddagger$ Protein Physiology Lab, Dep de Qu\'imica Biol\'ogica, Facultad de Ciencias Exactas y Naturales, UBA-CONICET-IQUIBICEN, Buenos Aires, Argentina \\
$^\S$ Laboratoire de physique statistique, CNRS, UPMC and \'Ecole normale sup\'erieure, 24 rue Lhomond, 75005 Paris, France\\
$^\star$ Laboratoire de physique th\'eorique, CNRS, UPMC and \'Ecole normale sup\'erieure, 24 rue Lhomond, 75005 Paris, France\\

$^\ast$ To whom correspondence should be addressed. Email: diegulise@gmail.com
}
\medskip
\end{minipage}

\begin{minipage}[c]{17cm}
\medskip
\small
\textbf{The analysis of correlations of amino acid occurrences in globular proteins has led to the development of statistical tools that can identify native contacts -- portions of the chains that come to close distance in folded structural ensembles. Here we introduce a statistical coupling analysis for repeat proteins -- natural systems for which the identification of domains remains challenging. We show that the inherent translational symmetry of repeat protein sequences introduces a strong bias in the pair correlations at precisely the length scale of the repeat-unit. Equalizing for this bias reveals true co-evolutionary signals from which local native-contacts can be identified. Importantly, parameter values obtained for all other interactions are not significantly affected by the equalization. We quantify the robustness of the procedure and assign confidence levels to the interactions, identifying the minimum number of sequences needed to extract evolutionary information in several repeat protein families. The overall procedure can be used to reconstruct the interactions at long distances, identifying the characteristics of the strongest couplings in each family, and can be applied to any system that appears translationally symmetric.}

\end{minipage}
\medskip

{\center{\rule[0mm]{175mm}{0.25mm}}}
\medskip


\begin{multicols}{2}

\noindent\large{\textbf{{Introduction}}}
\medskip
\normalsize
\par The fact that many protein molecules spontaneously collapse stretches of amino acid chains into defined structural domains \cite{pmid4351801} facilitates the description, evolution and construction of these peculiar physical objects. Higher order biological 
{\it functions} that are correlated with domains can usually be isolated, recombined and adjusted,
akin to engineering \cite{pmid18049465}, or tinkering \cite{pmid860134} using modular components. The evolutionary record of natural proteins results from a balance between sequence exploration and constraints: conservation of function within a protein family imposes strong boundaries on sequence variation, sculpting the structural forms visited by members of a protein family . Amino acids that are in spatial proximity in the mean conformational ensemble are expected to co-vary on evolutionary timescales, as their energy contributions to fold stabilization are often localized to groups of residues \cite{pmid9348663}. However, correlated residue changes throughout proteins' history may not necessarily be close in space, as other constraints are always at play \cite{pmid18077414}. Since the evolutionary record is inevitably incomplete, the sequences we find today constitute a biased sample of the possible outcomes, therefore any search for the underlying constraints must take into account contingent factors that may confound the observed correlations. Here we use sequence correlations to explore the link between structure and function in repeat proteins, natural systems for which the identification of functional domains remains challenging \cite{parra2013detecting}.

\begin{figure*}
\centering
\includegraphics{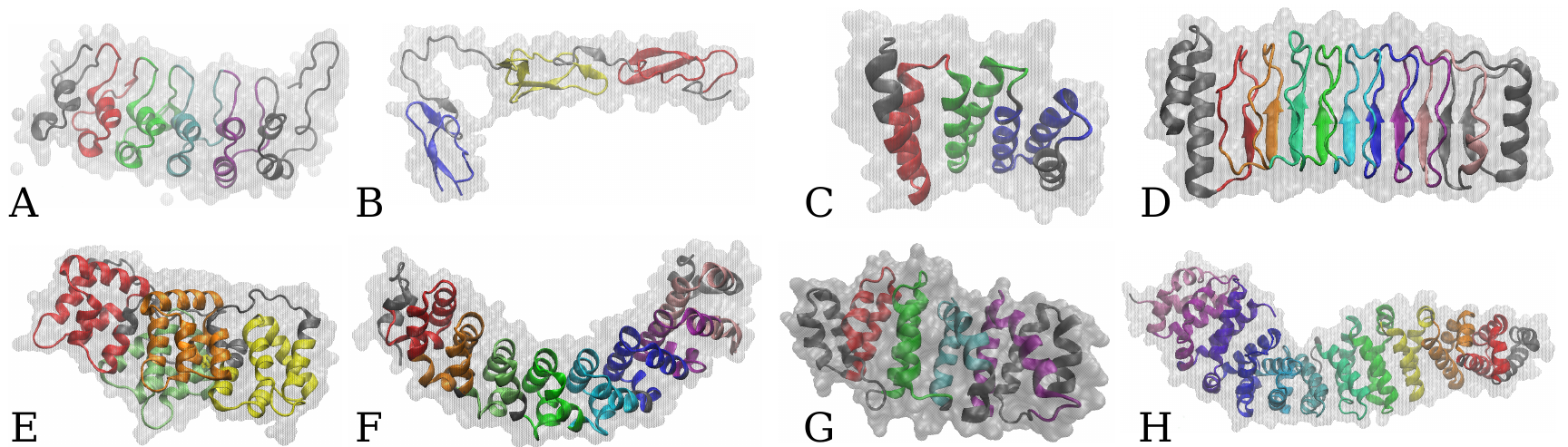}
\caption{Repeat proteins are formed with tandem arrays of repeats. The crystal structures of members of different repeat protein families are shown, with the backbone colored according to the repeated units. The molecular surface of the repeat array is drawn in transparent gray. \textbf{A.} \textsc{ank} family (PDB:1IKN, chain D), \textbf{B.} \textsc{egf} family (PDB:4D90, chain B), \textbf{C.} \textsc{tpr} family (PDB:4GCO), \textbf{D.} \textsc{leu} family (PDB:4NKH, chain A), \textbf{E.} \textsc{anex} family (PDB:2ZOC, chain A), \textbf{F.} \textsc{pum} family (PDB:2YJY, chain A), \textbf{G.} \textsc{heat} family (PDB:4G3A, chain A), and \textbf{H.} \textsc{arm} family (PDB:2BCT).}
\label{fig:structures}
\end{figure*}

\par Many natural proteins contain tandem repeats of similar amino acid stretches. These have been broadly classified in groups according to the length of the minimal repeating units \cite{kajava2012tandem}: short repeats up to five residues usually fold into fibrillar structures such as collagen or silk, while repeats longer than about 60 residues usually fold as independent globular domains.  There is a class of proteins whose repeat frequency lies in between these values and for which the folding of the repeating units is coupled. In these periodic repeat proteins unique ``domains'' are not trivial to define \cite{parra2013detecting}. Typical repeat proteins are made up of tandem arrays of $\sim$20-40 similar amino acid stretches that fold into elongated architectures of stacked repeating structural motifs (Fig.~\textcolor{blue}{\ref{fig:structures}}). Successful design of repeat proteins with novel functions based on simple sequence statistics \cite{tamaskovic2012designed} suggests that folding and functional signals can be partially segregated. Energy landscape theory predicts that foldable polypeptides are much easier to realize in the presence of symmetry as compared to asymmetric arrangements \cite{wolynes1996symmetry}. Funneled energy landscapes imply that patterns can form in different parts of the molecule with relative independence and subsequently assemble to higher order structures. This greatly reduces the folding search problem by efficiently arranging relatively small fundamental building blocks in a repetitive fashion \cite{ferreiro2008energy}. Thus, due to the approximate translational symmetry, repeat proteins constitute excellent systems in which to study the coupling between sequential, structural and functional patterns.

The maximum entropy principle proposes a scheme for approaching the problem of extracting essential pair couplings from multiple sequence alignments of families of homologous proteins \cite{neher1994frequent,weigt2009identification, mora2010maximum}. The main technical limitations confounding residue correlations are the transitivity of the correlations, the statistical noise due to the relative small number of available observables, and the phylogenetic dependence of the set of sequences assembled into a protein family  \cite{pmid24573474}. Indirect interactions may generate the dominant correlations, and disentangling direct from indirect links is a fundamental step towards inferring the energetics underlying the observed couplings \cite{weigt2009identification}. The application of statistical coupling analysis provides an efficient way of extracting meaningful information from the apparent junk of massive genomic data \cite{pmid9501075}. The mean structure of several protein domains can be reasonably well predicted from the statistical analysis of variations in large sets of sequences \cite{pmid22106262,pmid22691493}. Strong deviations of the statistically coupled positions from the known domain structures leads to explore dynamical aspects of proteins that are related to biological function \cite{pmid24297889}. Likewise, specific interactions between domains can be characterized and good approximations to the interaction energetics can be obtained \cite{marks2011protein,pmid24449878,pmid24160546, mora2010maximum}). Here we show the limitations of the statistical coupling analysis developed for globular proteins and propose an analogous procedure for quasi-translationally symmetric repeat proteins.

Specifically, we compare the information extracted from two-point correlation functions of multiple alignment of repeat protein sequences to the known structural interactions between the apparent repeated units. We show that the translational symmetry introduces a strong bias in the pair correlations at precisely the length scale of the repeated unit. Equalizing for this bias in an objective way results in correlation matrices from which local native-contacts can be identified. We apply this procedure to many families of repeat protein (introduced in Fig.~{\color{blue}\ref{fig:structures}}) and show that some families have strong interactions mainly between repeats, while others mainly within single repeats. These observations can be linked to the functional characteristics of the families.

\medskip
\noindent\large{\textbf{{Results}}}
\medskip

\small{\textbf{Direct coupling analysis of repeat proteins}}
\normalsize
\par To characterize correlations between amino acid positions in natural repeat proteins we needed to define a length scale on which to search, align and compose a repeat region. We chose to use the minimal definition of repeats present in the \textsc{pfam} database \cite{bateman2004pfam}, and obtained sequences of single repeated units for the families listed in Table 1 of SI. Since a repeat domain is formed with multiple tandem copies of repeats units \cite{parra2013detecting}, the minimal sequence that includes an interface between repeats is composed of two consecutive units. We thus constructed multiple sequence alignments (\textsc{msa}) of pairs of consecutive repeats for each family. The sets of sequences were corrected for phylogenetic bias and finite-size sampling as described in the Methods. 

\begin{figure*}
\centering
	\includegraphics{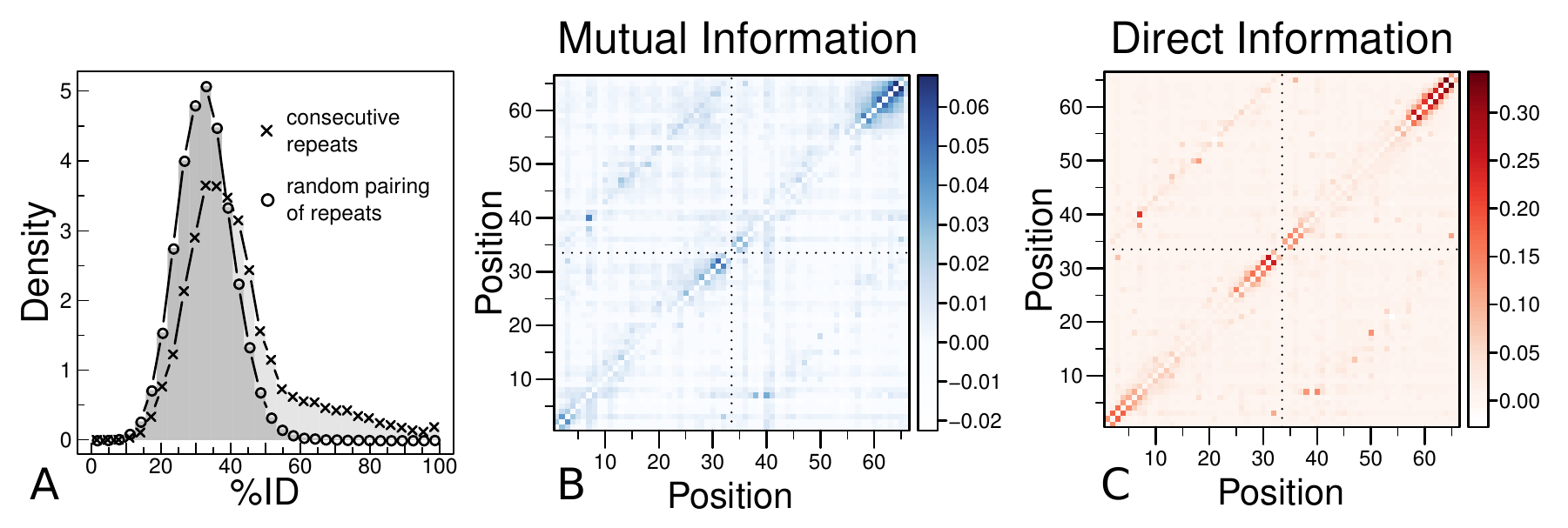}
	\caption{The sequence identity between repeated units can bias the inference of evolutionary couplings. Repeat sequences of the \textsc{ank} family were concatenated in a MSA of size $2L_0=66$ positions and $\approx$73000 sequences and co-variations were measured with mutual and direct information metrics. \textbf{A.} Sequence identity distributions between consecutive \textsc{ank} repeats found in (x) natural proteins and (o) randomized pairs of repeats. \textbf{B.} Mutual information and \textbf{C.} Direct information matrices between positions obtained without correcting (upper half) or with proper equalization for repeat identity (lower half) .}
	\label{fig:dcaresults}
\end{figure*}

Mutual information (\textsc{mi}) and direct information (\textsc{di}) use covariance in homologous protein sequences to deduce structural constraints. While \textsc{mi} uses the joint frequencies of aminoacids (eq. \ref{eq:MI}), \textsc{di} (eq \ref{eq:DI}) uncouples direct interactions from interactions mediated by a third residue on the complete sequence of the protein.
The upper triangles of figures {\color{blue}\ref{fig:dcaresults}B} and {\color{blue}\ref{fig:dcaresults}C} show the \textsc{mi} and \textsc{di} matrices for one of the most abundant repeat proteins, the Ankyrin-repeat family. The typical length of these repeats is 33 residues, so values on columns/rows 1 to 33 and 34 to 66 correspond to interactions between residues within a repeat, while values on columns 1 to 33 and rows 34 to 66 correspond to interactions between residues on consecutive repeats.  
Both \textsc{mi} and \textsc{di} present overall similar patterns with \textsc{mi} having a noisier background signal. The values corresponding to pairs of positions on consecutive repeats reach comparable values to those within each repeated unit. There appears to be as much evolutionary correlations between residues on the same repeat as between residues in consecutive repeats. A question that arises is whether the strong signal between repeats is due to the inevitable similarity of the sequences of repeat regions or to true coevolutionary interactions between neighboring repeats.

\par A close inspection of the couplings detected between repeated units reveals that the strongest signals are attributed to pairs of positions that are 33 residues apart (Fig. {\color{blue}\ref{fig:dcaresults}B} and {\color{blue}\ref{fig:dcaresults}C}, upper triangle). Since the ankyrin repeats aligned are of this precise length $L_0$, these apparent interactions occur between residues that occupy equivalent positions in each repeat, i.e:  the pair of positions ($i, i+L_0$) corresponds to the $i$th residue on the first repeat and the $i$th residue on the second repeat. If repeats in proteins were identical, the interactions between residue $i$ and $i+L_0$ should get maximum  \textsc{mi} and \textsc{di} values as these would show perfect co-variation. At the same time, the submatrix of positions between repeats should be identical to the submatrix of pairs of positions within the repeats. Thus, the identity between repeated units should be taken into account when evaluating correlations between repeats.

\par To characterize how the identity between neighboring repeats affects the covariation analysis, we compared the distribution of the percentage of identical residues, $\%Id$, between pairs of consecutive repeats, and between randomly assembled pairs of repeats (Fig. \textcolor{blue}{\ref{fig:dcaresults}}). For the ankyrin family, the distribution of  $\%Id$ for random pairs is a Gaussian centered around 30\%, while the natural pairs show higher mean and a large tail towards higher $\%Id$ values (Fig. \textcolor{blue}{\ref{fig:dcaresults}A}). This higher similarity between pairs of consecutive repeats is expected to induce correlations bewteen $i$ and $i+L_0$ positions, as observed. To compensate for the higher $\%Id$ between natural repeats we developed a correction factor that equalizes the effects of quasi-translational symmetry. This correction consists of calibrating the weight of each sequence in the natural neighbors according to the $\%Id$ between the component pair of repeats, and rescaling it so that it matches the expected frequency of $\%Id$ between random pairs of repeats of the same family (see Methods). We refer to the obtained values as DI$_{id}$ and MI$_{id}$. Figures \textcolor{blue}{\ref{fig:dcaresults}B} and \textcolor{blue}{\ref{fig:dcaresults}C} show the  \textsc{di} and  \textsc{mi} corrected only for phylogeny and finite counts (upper triangles), together with the ones that include this additional factor DI$_{id}$ and MI$_{id}$ (lower triangles). The strong symmetric ($i, i+33$) off-diagonal signal is attenuated for both  \textsc{mi} and  \textsc{di}, as expected if the signal originates from biases in the $\%Id$ distributions. Importantly, the  \textsc{di} values obtained for interactions between all other positions are not significantly affected by the $\%Id$ equalization.

\par The same analysis was performed for all the other repeat protein families (see Suppl. material, Fig. \textcolor{red}{\ref{fig:S1}}). The results for the \textsc{tpr}, \textsc{egf} and \textsc{leucine rich} families show a strong bias in the symmetric ($i, i+L_0$) interactions. These also show a higher sequence identity between true first neighboring repeats, which biases the inter-repeat couplings. Families \textsc{arm}, \textsc{spectrin}, \textsc{annexin} and \textsc{pumilio} do not show a high ($i, i+L_0$) signal on the \textsc{di} and \textsc{mi} matrices. For these, the distributions of $\%Id$ between true and random neighbors are similar, consistent with the notion that the symmetric signal is caused just by the bias in similarity between neighboring repeats. Applying the  $\%Id$ equalization to these families does not significantly change the \textsc{di} and \textsc{mi} values, showing that the correction is not detrimental to the overall procedure. Finally, the NEBULIN family does not present a strong ($i, i+L_0$) signal, and the \textsc{heat} family has a very rugged $\%Id$ distribution. We believe that these effects are caused by an insufficient number of effective sequences on the alignments, which cannot ensure a robust calculation of \textsc{di} and \textsc{mi} ({\it vide infra}). We conclude that sequences of proteins that show quasi-translational symmetry should be treated with an additional correction factor to account for the biases that the internal sequence identity can bring about.

\medskip
\small{\textbf{Prediction of native contacts for repeat proteins}}
\normalsize
\par For several globular domains it has been shown that native contacts can be inferred from the inspection of the top-list of residue pairs according to the \textsc{di} ranking \cite{pmid24573474}. There is no established way to discern the minimum value of \textsc{di} to be used as the cutoff, as these depend on the topology of the fold, the sampling of sequences and the details of the method used to obtain \textsc{di}, thus 50 to 200 pairs are empirically used. Since domains of repeat proteins are composed with multiple copies of repeated units, we asked whether \textsc{di} and DI$_{id}$ metrics are useful predictors of direct native interactions at the sub-domain level. We observed that the absolute values of \textsc{di} we calculated for pairs of repeats are lower than those computed for globular domains, (Fig. \textcolor{blue}{\ref{fig:dcaresults}} and \textcolor{red}{\ref{fig:S1}}), complicating the distinction of positive \textsc{di} outliers from the background signal. We developed a clustering method to objectively delimit the true positive interactions. We first calculated the euclidean distance between each pair of \textsc{di} values as $dDI_{a,b} = \sqrt{(DI_a - DI_b)^2}$; and made a hierarchical clustering of the obtained distances. To delimit the clusters we used the dynamic tree cut method \cite{langfelder2008defining}, which allows us to distinguish nested clusters. We found that most of the \textsc{di} pairs fall in one big cluster which we assigned to the background signal (Fig. \textcolor{red}{\ref{fig:S2}}). The other clusters have fewer members and constitute outliers of the normal distribution. We consider the true coevolutionary signals as those within small clusters of positive \textsc{di} values.
\par
\begin{figure*}
\centering
	\includegraphics{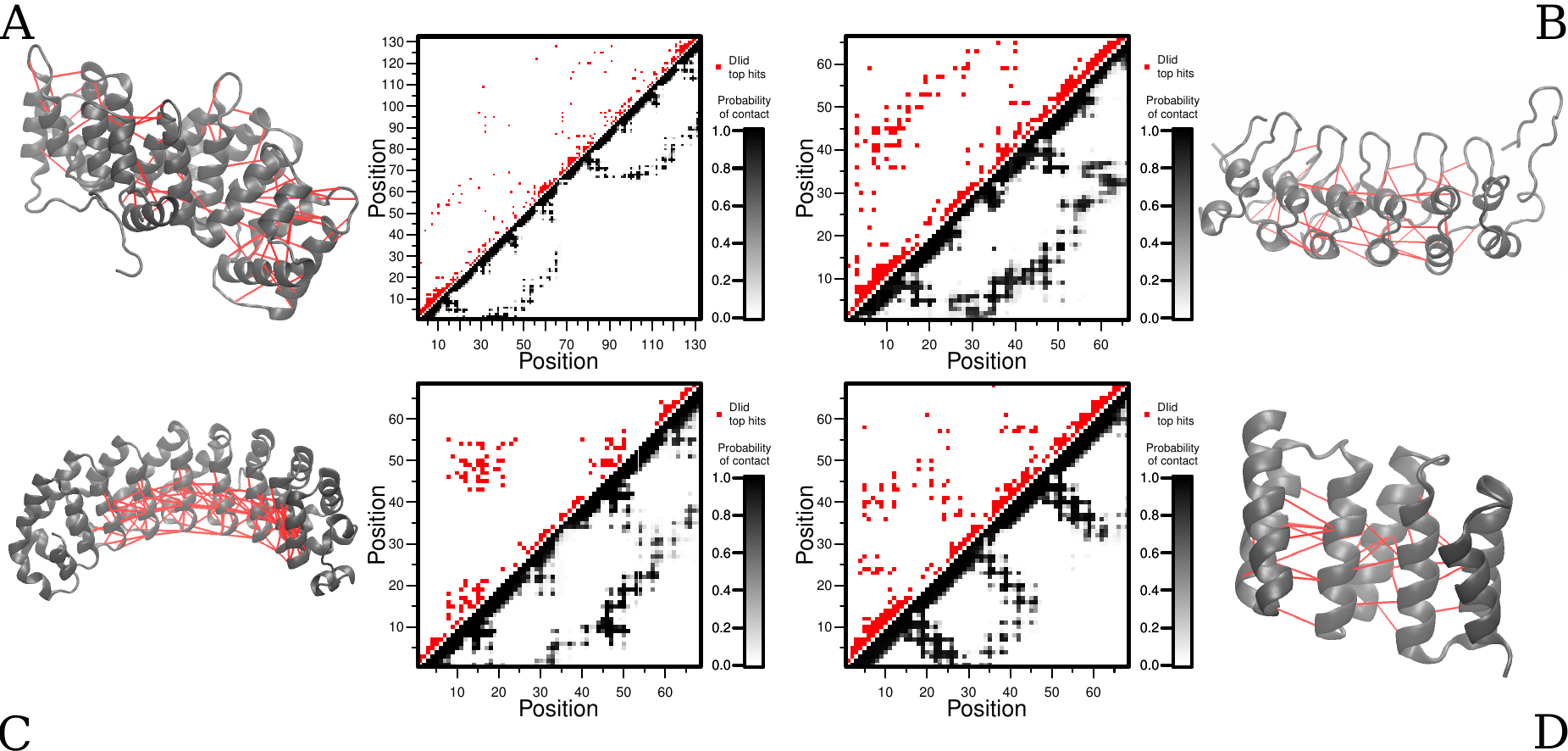}
	\caption{Native contacts can be predicted from the identity-equalized direct information DI$_{id}$. The probability of contact formation along ensembles of structures of consecutive repeat pairs in several repeat protein families is shown (lower triangles), together with the contact prediction based on the DI$_{id}$ distribution (upper triangles). The structure of representative family members are shown on the sides, with the backbones as gray ribbons and the first 20 predicted contacts along multiple repeat pairs in red. \textbf{A.} \textsc{anex} (PDB:2ZOC, chain A) \textbf{B.} \textsc{ank} (PDB:1IKN, chain D) \textbf{C.} \textsc{pum} (PDB:2YJY, chain A) \textbf{D.} \textsc{tpr} (PDB:4GCO)}
	\label{fig:DCAandStructures}
\end{figure*}

 Several high-resolution structures for repeat proteins are available. These typically fold into elongated architectures where most members of a family display an overall similar topology (Fig. \textcolor{blue}{\ref{fig:structures}}). Notably, the repeat arrays can vary in the number of repeat units and many details and irregularities plague the structural representatives \cite{parra2013detecting}. To get an overall representation of the distribution of contacts in the known structures we computed the probability of contact formation along the ensembles of structures as described in Methods. We obtained an average contact map of the repeat architecture mapped on to the sequences of the \textsc{msa}, where residue pairs with high density correspond to residue pairs that are most frequently encountered within contact distance (Fig. \textcolor{blue}{\ref{fig:DCAandStructures}}, lower triangle). The pattern of evolutionary interactions inferred from the clustering of DI$_{id}$ is remarkably similar to the experimental contact map densities for most families (Fig. \textcolor{blue}{\ref{fig:DCAandStructures}} and \textcolor{red}{\ref{fig:S2}}). The signals from the pairs of positions of consecutive repeats ($i, i+L_0$) do not always correspond to a high contact probability, yet if present they are confidently detected.  
\par One of the longest pairs of repeated units we study belongs to the Annexin family ($2L_0\approx$ 132 residues). The DI$_{id}$ hits strongly resemble the average contact map, with 113 out of the 150 DI$_{id}$ pairs found within contact distance in at least half of the experimental structures (Fig \textcolor{blue}{\ref{fig:DCAandStructures}A}). 
Most of these correspond to interactions within each repeat, with few interactions at the repeat interfaces, unlike the correlations found in other repeat proteins, such as the Ankyrin family (Fig. \textcolor{blue}{\ref{fig:DCAandStructures}B})
\par The clustering procedure assigns 246 hits for DI$_{id}$, 166 of which are typically found within contact distance. Most of these are found outside the usual binding site of these proteins -- the $\beta$-hairpin motif.
Coevolutionary interactions of the Pumilio family also map to positions between repeated units, with 113 experimental contacts out of 150 predicted (Fig. \textcolor{blue}{\ref{fig:DCAandStructures}C}). Yet in this case they are mainly clustered around the regions where these proteins bind nucleic acids. Within the top 237 DI$_{id}$ identified for the Tetratricopeptide family, only 167 are typically found within contact distance in the experimental structures and most of the outliers are in regions physically compatible with the known structures (Fig. \textcolor{blue}{\ref{fig:DCAandStructures}D}). A similar picture is apparent in the Armadillo family, where only 140 interactions correspond to mean contacts among the 231 predicted (Fig. \textcolor{red}{\ref{fig:S2}}). In the case of the Leucine-rich family, few interactions appear as outliers in DI$_{id}$ distribution, and most of them have been observed to form close contacts between repeated units (Fig. \textcolor{red}{\ref{fig:S2}}). Repeats of the \textsc{egf} family rarely interact, and DI$_{id}$ consistently fails to detect inter-repeat correlations, acting as a negative control for the overall procedure (Fig. \textcolor{red}{\ref{fig:S2}}). Finally, few co-evolutionary interactions are assigned in the \textsc{heat} family, probably due to the limited number of available sequences (see below). Since there are no experimental structures for the Nebulin family, we cannot evaluate if the identified DI$_{id}$ hits correspond with native-contacts.

\medskip
\medskip

\medskip
\small{\textbf{Distant couplings along a repeat array}}
\normalsize
\par Folding of repeat domains usually involves the cooperative formation of structures at a length scale that exceeds first neighbors \cite{aksel2009analysis}. Folding in some regions nucleates the folding of contiguous segments, allowing for a quasi-one-dimensional treatment of the dynamics \cite{ferreiro2008capillarity}. A natural question that arises is how do evolutionary couplings in and between repeats change as the separation between the repeats increases.

\hspace*{3pt}
\medskip
\begin{figurehere}
\centering
		\includegraphics{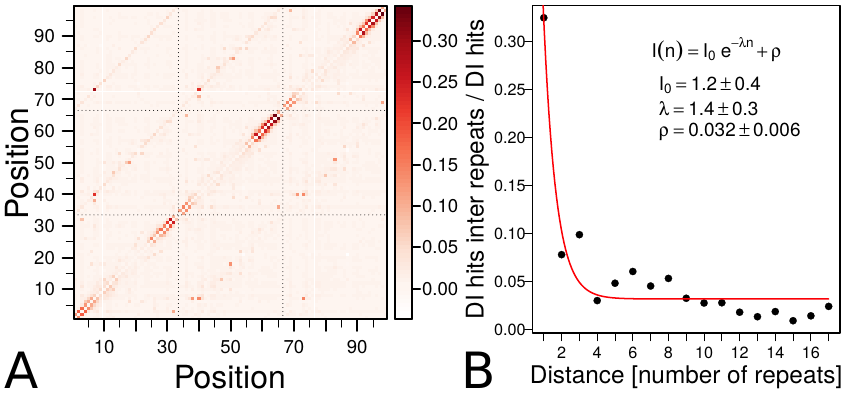}
		\caption{Correlations along \textsc{ank} repeat arrays. \textbf{A.} Direct information matrix calculated for three pairs of \textsc{ank} repeats without (upper triangle) or with (lower triangle) the DI$_{id}$ equalization. \textbf{B.} Proportion of DI$_{id}$ hits between repeated units for alignments of $n$-th neighbors. The line is a non-linear fit of the data to an exponential decay. }
		\label{fig:moreneighbours}
\end{figurehere}
\hspace*{3pt}

\par An analogous correction to the weights of the sequences must be made to treat $n$-neighbors interactions (see lower triangle of Fig.~\textcolor{red}{\ref{fig:S3}} for the uncorrected DCA of three consecutive repeats of the ankyrin family). When the proper equalization is performed, the symmetric signals attenuate and the true coevolutionary correlations appear (DI$_{id}$ lower triangle of Fig.~\textcolor{red}{\ref{fig:S3}}). In principle the correction to the symmetric ($i, i+n L_0$) interactions can be applied to arbitrarily large repeat proteins. Yet the sampling needed is much larger and the computing time growths as L$^2$, restricting the application to longer repeat arrays. Since in \textsc{ank}s, as in most of the repeat protein families, interactions are concentrated at relatively short sequences separations, we reconstructed a DI$_{id}$ matrix from a parallel calculation of repeat pairs. For first neighbors we estimated DI$_{id}$ as described previously, and for second neighbors we concatenated the sequences in an \textsc{msa} of size $2L_0$. The reconstructed matrix for all interactions is very similar to the one calculated on the whole three-repeat \textsc{msa} (Fig.  \textcolor{blue}{\ref{fig:moreneighbours}A} and {\color{red}\ref{fig:S3}}), facilitating the application of the analysis for larger repeat arrays. 
\par We observed that as the separation between repeats increases, the DI$_{id}$ between repeats decays significantly (Fig. \textcolor{blue}{\ref{fig:moreneighbours}B}). True repeat pair interactions are less frequent, and this is reflected in the evolutionary couplings between units.
The number of interactions between repeats decreases roughly exponentially with repeat separation, with a half-length of about 1.4 repeats (Fig. \textcolor{blue}{\ref{fig:moreneighbours}B}), suggesting that the evolutionary correlation length of Ankyrin repeat arrays is $\sim$1.5 units.

\medskip
\small{\textbf{Robustness and confidence of the analysis}}
\normalsize
\par For a robust calculation of the \textsc{di} one must have a sufficiently large number of effective sequences to approximate the marginal and joint probability distributions from the observed frequencies of occurrences of amino acids. Since there is no general principle indicating how many sequences are necessary and sufficient for robust estimation, we empirically quantified the minimum number of effective sequences in various repeat protein families.

\hspace*{3pt}
\medskip
\begin{figurehere}
\centering
	\includegraphics{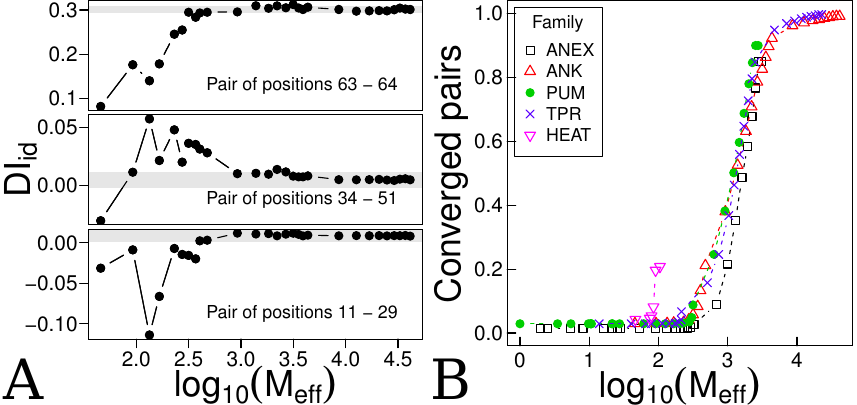}
	\caption{Robustness of the DI$_{id}$ procedure. Subsets of alignments were constructed by recurrently removing random groups of sequences from each dataset of repeat pairs. M$_{eff}$ is the number of effective sequences used in the alignment. \textbf{A.} Particular examples of the stability of DI$_{id}$ assignments as sampling changes on the \textsc{ank} family. The gray shadow delimits the 1\% fluctuation interval set as a convergence criteria. \textbf{B.} Overall stability of the DI$_{id}$ assignments in several repeat protein families.}
\label{fig:convergence}
\end{figurehere}
\hspace*{3pt}

\par We constructed subsets of alignments by recurrently removing random groups of sequences from each dataset of repeat pairs, and calculated \textsc{dca} on each of these subsets. The reduction in the number of sequences typically decrease the absolute values of the high ranking DI$_{id}$ matrix elements and at the same time increases the background DI$_{id}$ signals (Fig. \textcolor{red}{\ref{fig:S4}}), making both MI$_{id}$ and DI$_{id}$ signals indistinguishable from the background for small sample sizes.

\par For well determined parameters we expect the true value will be better estimated as sampling increases. Examples of the robustness of the DI$_{id}$ assignments are shown in the panels of Fig.~\textcolor{blue}{\ref{fig:convergence}A}. While the DI$_{id}$ of some residue pairs can be confidently established with about 500 effective sequences, other pairs do not reach stable values even when all the available sequences are taken into account (Fig.~\textcolor{blue}{\ref{fig:convergence}A}). To globally quantify the convergence of the DI$_{id}$ matrix we evaluated how many of the residue pairs reach a limiting value within 1\% of the one obtained with the largest sample size. For every subset of sequences, $s$, we require that $|DI_{ij}^{s}-DI_{ij}|<0.01 \cdot \left(\max(DI)-\min(DI)\right)$, where $DI_{ij}^{s}$ is the \textsc{di} between position $i$ and $j$ calculated over the $s$-th subset, $DI_{ij}$ is the \textsc{di} on the largest set of sequences, and $\max(DI)$ and $\min(DI)$ are the maximum and minimum values for all positions in all subsets. Additionally all subsets larger than the subset $s$ one must have a standard deviation lower than 1\% of the standard deviation of the \textsc{di} values from all the subsets. If a residue pair fulfills these conditions, we say it has converged at the particular $s$ sample size. We quantified how many of the residue pairs satisfy the convergence criteria at various sample sizes (Fig.  \textcolor{blue}{\ref{fig:convergence}B}). The best sampled families, \textsc{ank} and \textsc{tpr}, contain enough sequences to converge the DI$_{id}$ for almost all residue pairs of consecutive repeats. Reducing the number of input sequences results in a loss of convergence of some sites; the DI$_{id}$ of around 90\% of the residue pairs can be confidently established with about 10\% of the total sequences ($M_{eff}\approx10^{7/2}$) (Fig.  \textcolor{blue}{\ref{fig:convergence}B}). If the subsamples are further reduced, the proportion of positions that converge drops catastrophically. Yet even more relaxed criteria for convergence give confident results for the high-ranking \textsc{di} pairs, as exemplified by the \textsc{pum} and \textsc{anex} families (Fig.  \textcolor{blue}{\ref{fig:convergence}B}). However the samples for the \textsc{heat} family are not sufficient to confidently quantify repeat pairs co-evolution.

\medskip
\noindent\large{\textbf{{Discussion}}}
\normalsize
\par Repeat proteins are formed with various tandem repetitions of similar amino acid stretches. Due to the approximate translational symmetry, regions in proximity in the amino acid chain show similarities in their sequence patterns, which can result in close to perfect co-variation in a multiple sequence alignment and hence bias the inferred interactions between residues (Fig. \textcolor{blue}{\ref{fig:dcaresults}}). To compensate for this natural bias we developed an equalization that re-weighs each sequence in the multiple alignment to account for correlations characteristic of the protein family. This procedure reveals the true co-evolutionary signals in the case of strong biases, importantly leaving the quantifications unchanged in the absence of bias.

The DI$_{id}$ metric resulting from this corrected statistical coupling analysis is a good predictor of native interactions at the sub-domain level for proteins with a quasi translational symmetry, similarly to the original \textsc{di} metric for globular proteins \cite{pmid24573474}. The highest ranking DI$_{id}$ pairs are usually found in spatial proximity in all of the repeat protein families analyzed (Figs. \textcolor{blue}{\ref{fig:DCAandStructures}} and \textcolor{red}{\ref{fig:S2}}). Interestingly, the patterns of co-evolutionary interactions are not a random subset of all the native-interactions, but segregate into particular groups in each family. Some families display relative high inter-repeat correlations, while in others the repeats appear to be independent evolutionary units. In their native environment, most repeat proteins participate in binding other macromolecules, and are thus expected to show co-variations in the positions that correspond to the binding interface. We observed that some architectures do show higher co-variations at the typical binding interface, like the nucleic-acid binding \textsc{pum} family, while in the ubiquitous \textsc{ank} family the typical binding interface is depleted of DI$_{id}$ pairs.

\par  A reliable estimation of \textsc{di} requires a sufficiently large number of
sequences. This number depends on the length, the topology and the ontology of the proteins under scrutiny. We empirically quantified the minimum number of effective sequences needed by sampling the subsamples of repeat protein families (Fig. \textcolor{blue}{\ref{fig:convergence}}). In most families we found that $\sim$90\% of the residue pairs can be confidently established with $\sim10^{7/2}$ sequences (Fig. \textcolor{blue}{\ref{fig:convergence}}). The highest ranking \textsc{di} interactions confidently predict native contacts even for much scarcer sampling.

\par Repeat proteins usually fold cooperatively several consecutive repeats \cite{aksel2009analysis}. Nucleation of the folding in some region facilitates the folding of contiguous segments, allowing for a quasi-one-dimensional treatment of the dynamics \cite{ferreiro2008capillarity}. We found that the statistical couplings calculated from sequence variations in the \textsc{ank} family decay roughly exponentially (Fig. \textcolor{blue}{\ref{fig:moreneighbours}}) as the separation between the repeats increases. The predicted global correlation length of $\sim$1.4 repeated units is remarkably close to that inferred from statistical mechanical analysis of folding experiments \cite{street2009predicting,wetzel2008folding} and folding simulations \cite{ferreiro2005energy}. These predictions are based on approximating long-range covariations from sets of pair-wise inter-repeat interactions, allowing for the application of the procedure for arbitrarily large structures for which an exact calculation would be computationally prohibitive.


\medskip
\noindent\large{\textbf{{Materials and methods}}}

\small{\textbf{Alignments data}}
We obtained the \textsc{msa} for repeat units with \textsc{ncbi} data from the \textsc{pfam} \cite{finn2013pfam} database for the families listed on Table 1 of SI. For each \textsc{msa} we ignored the columns that contain gaps in more than the 80\% of the members. The remaining number of residues in each case is referred as $L_0$. In order to reconstruct tandem arrays of repeats, we concatenated the sequences that belong to the same protein (as identified in Uniprot \cite{TheUniProtConsortium01012014}), and for which the sequence separation is less than $L_0 /3$. The alignment thus generated is referred as first neighbor alignment and has $L=2L_0$ columns (positions) with $M$ rows (sequences) for each of the prototypical families of repeat proteins listed in Table 1 of SI.

\medskip
\small{\textbf{DCA calculations}}
 On every constructed \textsc{msa} we performed \textsc{dca} using the matrix inversion method detailed in \cite{pmid22106262}. To correct for the phylogenetic bias in the ensembles of sequences, we weighted them with the Henikoff and Henikoff heuristic \cite{henikoff1994position}, by assigning a weight 
$
w_i=\sum_j \frac{1}{r_j\cdot s_j^ i}
$
to each sequence. $r_j$ is the number of different amino acids present in position $j$ of the MSA and $s_j^i$ is the number of sequences that have the same amino acid on position $j$ than sequence $i$. We approximated the effective number of sequences as $M_{eff}=\sum_i w_i$. We calculated the mutual information (\textsc{mi}) and direct information (\textsc{di}) as: 
\begin{eqnarray}
MI_{ij}=\sum_{A,B}f_{ij}(A,B)\ln\left(\frac{f_{ij}(A,B)}{f_i(A)f_j(B)}\right) \label{eq:MI}\\
DI_{ij}=\sum_{A,B}P^{dir}_{ij}(A,B)\ln\left(\frac{P^{dir}_{ij}(A,B)}{f_i(A)f_j(B)}\right) \label{eq:DI}
\end{eqnarray}
where $f_i(A)$ is the marginal frequency of amino acid $A$ at position $i$ of the  \textsc{msa}, $f_j(B)$ is the marginal frequency of amino acid B at position $j$ of the  \textsc{msa}, $f_{ij}(A,B)$ is the joint frequency of having amino acid $A$ at position $i$ and amino acid $B$ at position $j$ simultaneously and $P^{dir}_{ij}(A,B)$ is the probability of having amino acid $A$ at position $i$ and amino acid $B$ at position $j$ simultaneously generated by the direct coupling between these pairs of residues.
\par The finite-size of the ensemble of sequences generates spurious correlations that must be corrected for. By scrambling each of the columns of a natural \textsc{msa} we generate MSA$_{IM}$ which keeps the marginal frequencies of the amino acids in each position but breaks all true correlations. We calculated mutual and direct information for this site-independent alignment and subtracted the results from the mutual and direct information calculated on the original \textsc{msa}. These values are presented in the matrices \textsc{mi} (for mutual information) and \textsc{di} (for direct information).

\medskip
\small{\textbf{DI$_{id}$ calculation}}
 We accounted for the self-similarity of repeats by weighting the sequences according to the sequence identity of a repeat pair. We calculated the percentage of identical residues $\%Id$ between the repeats on the same sequence ($\nu(\%Id)$) and the randomly expected $\%Id$ between M pairs of repeats of the same family, but belonging to different proteins, $\nu^{random}(\%Id)$. Since aligned repeats have L$_0$ residues each, the $\%Id$ can only take discrete values $n/L_0$ with $n$ an integer between 0 and L$_0$. We weighted each sequence by: \begin{equation}
w^{c}_i= w_i \frac{\nu^{random}(\%Id=n_0L_0)}{ \nu(\%Id=n_0L_0)}
\end{equation}
where $w_i$ is the Henikoff weight of a sequence that has $\%Id=n_0 L_0$. The \textsc{dca} calculations that include these weights are referred to as DI$_{id}$ and MI$_{id}$. 

\medskip
\small{\textbf{Density of contacts map}}
To compare the results of \textsc{di} and \textsc{mi} calculations with available structural models of repeat proteins we took all available structures from the protein data bank \cite{bernstein1977protein} cataloged under the \textsc{pfam} accession number of the family. 
The numeration of the residues for the repeat units were identified with HMMER \cite{finn2011hmmer} and the corresponding MSA.
We calculated a contact map for each PDB structure based on the euclidean distances between the C$\alpha$ atoms of each amino acids. We considered amino acids to be in contact if their C$\alpha$ are closer than 10 \AA. The contact probability of a pair of residues was defined as the number of times this pair is found in contact in the ensemble of structures.

\medskip
\noindent{\textbf{Acknowledgments}}
\small
Work was supported by the Consejo Nacional de Investigaciones Cient\'ificas y T\'ecnicas de Argentina (CONICET), the Agencia Nacional de Promoci\'on Cient\'fica y Tecnol\'ogica (ANPCyT), and ERCStG n. 306312. .

\bibliographystyle{apalike}
\bibliography{bibliography.bib}

\end{multicols}

\newpage

\setcounter{page}{1}
\setcounter{figure}{0}

\renewcommand{\thefigure}{S\arabic{figure}}
\renewcommand{\thepage}{S\arabic{page}}

\begin{center}
\textbf{\LARGE Supplementary material}\\
\end{center}

\medskip

\section*{MI and DI calculations over repeat protein families.}

\begin{figurehere}
\centering
 	\includegraphics{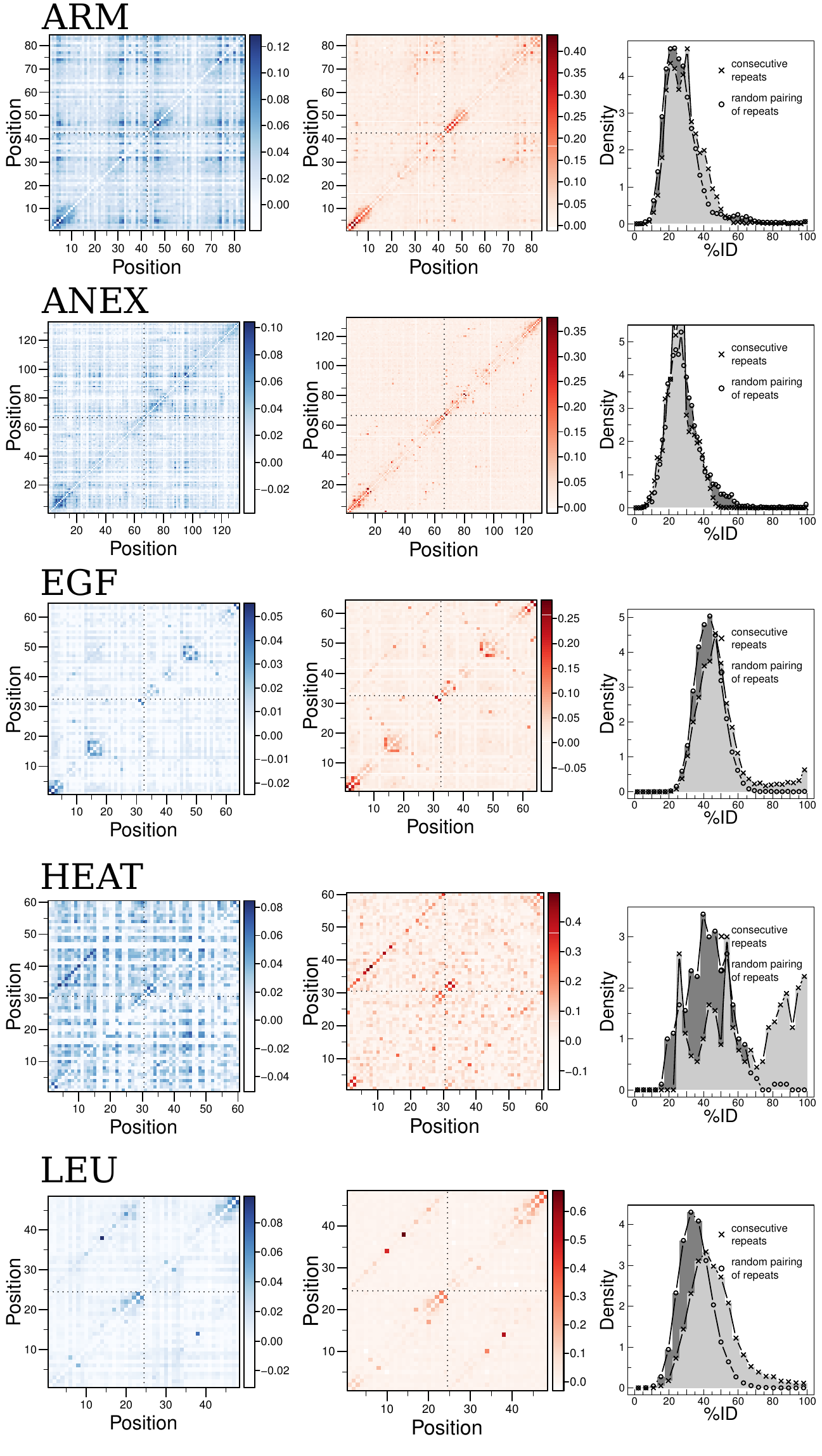}
\end{figurehere}

\begin{figurehere}
\centering
 \includegraphics{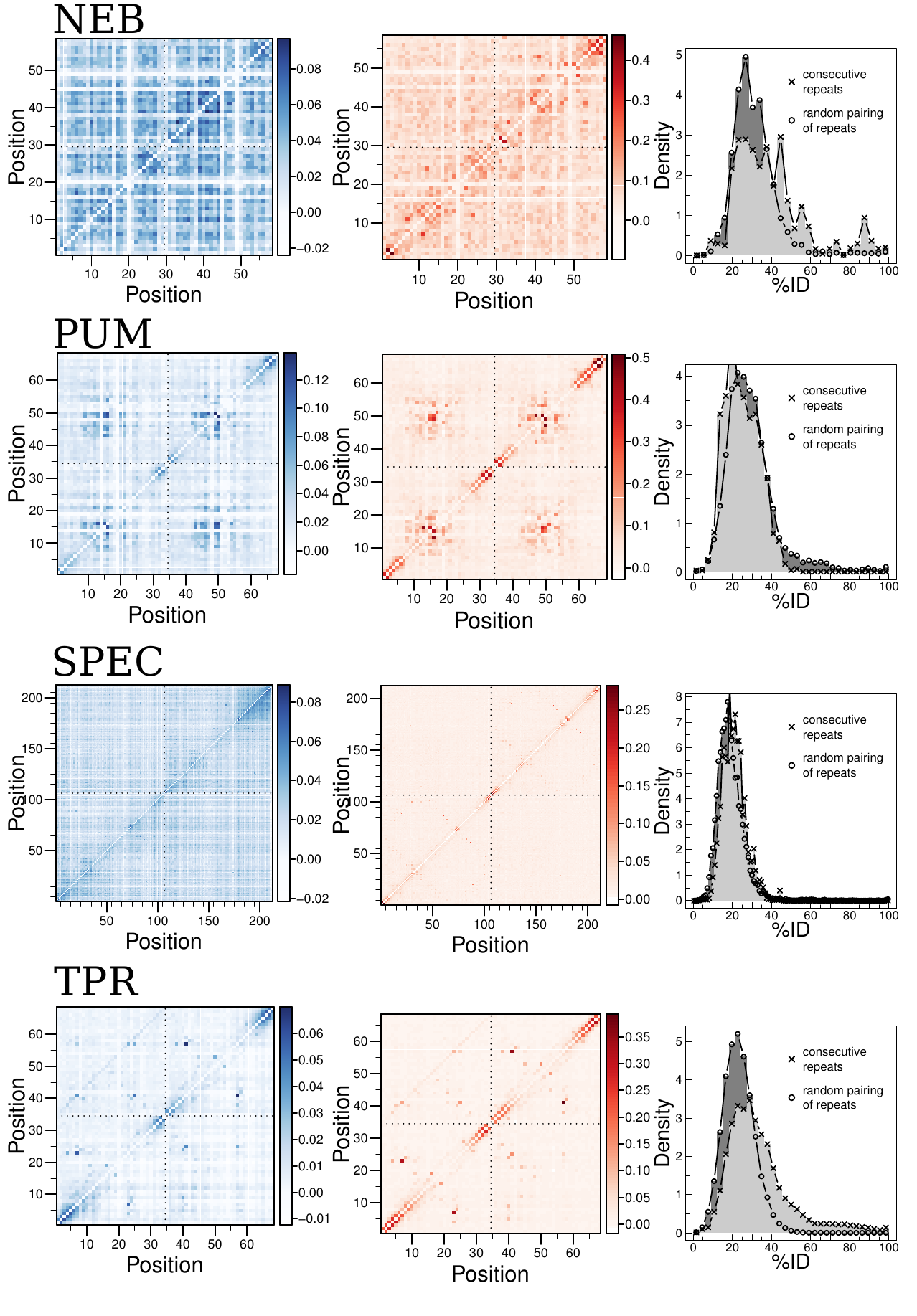}
 \caption{For each family, the blue matrix is \textsc{mi} on the upper triangle and MI$_{id}$ on the lower triangle; the red matrix is \textsc{di} on the upper triangle and DI$_{id}$ on the lower triangle; the third pannel has the comparison between histograms of \%Id for the FNA (first neighbours repeats - x) and the RPA (random pairs of repeats alignment - o).}
 \label{fig:S1}
\end{figurehere}

\section*{Selection of top DI hits.}

\begin{figurehere}
	\centering
 	\includegraphics{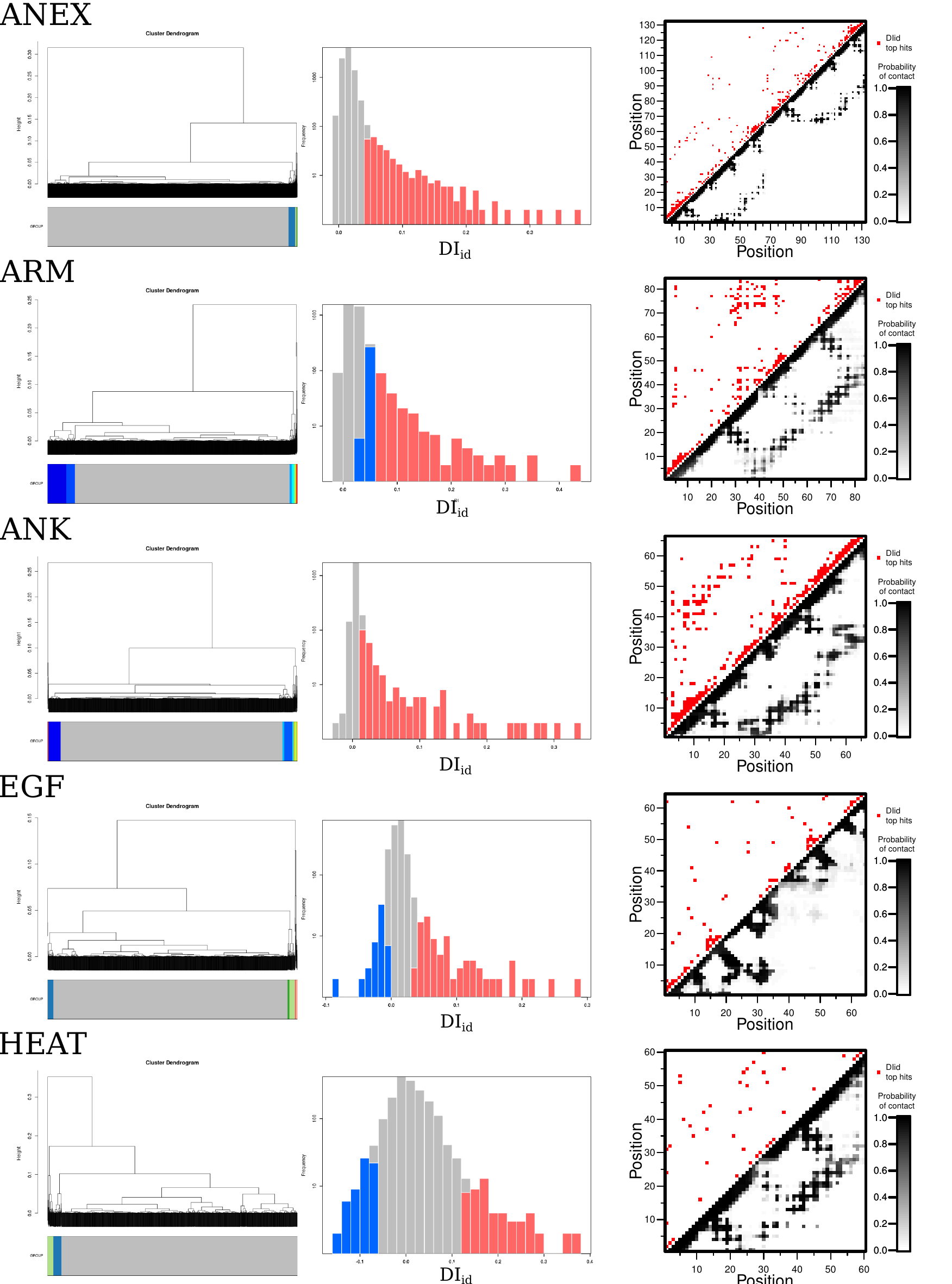}
\end{figurehere}

\begin{figurehere}
\centering
 \includegraphics{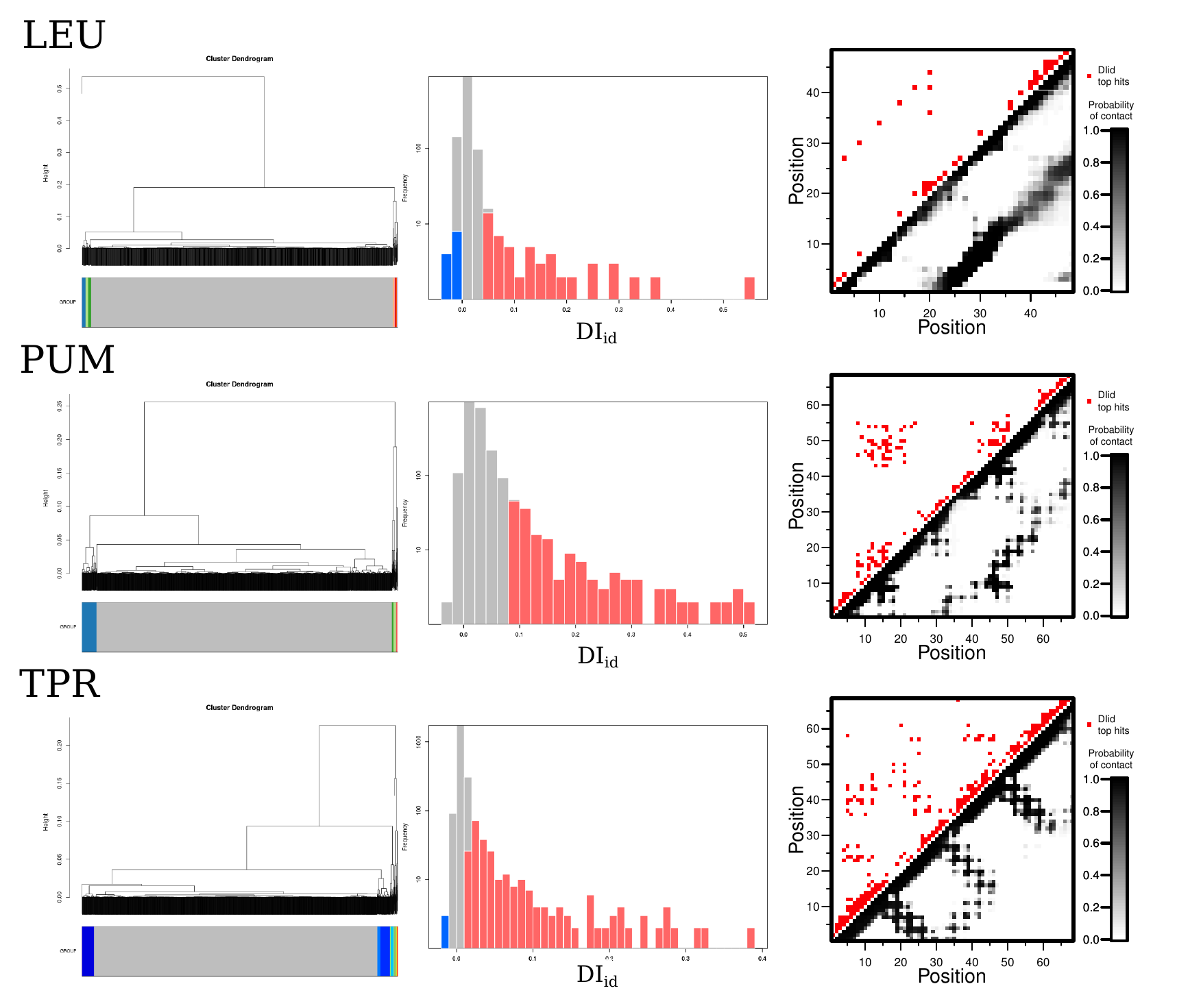}
 \caption{For each protein family, on the left the dendogram of dDI$_{id}$: each leave is a pair of positions and the height is the ablsolute value of difference of \textsc{di}. In the center the histogram of DI$_{id}$ values. On red and blue the small clusters distributions; on red ones considered DI$_{id}$ hits, and on blue the ones that were not. The third pannel, on top the DI$_{id}$ hits and below the probability of contact map. }
 \label{fig:S2}
\end{figurehere}

\section*{Distant couplings along a repeat-array}

\begin{figurehere}
\centering
 \includegraphics{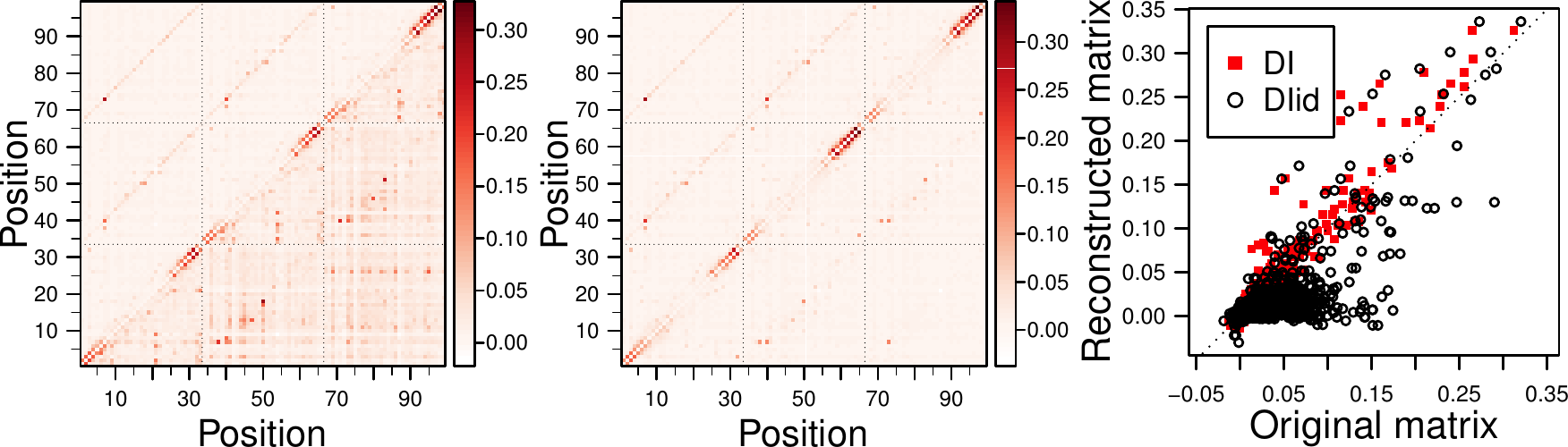}
 \caption{Left, upper triangle \textsc{di} and bottom triangle DI$_{id}$ for the three repeats alignment. Center, upper triangle \textsc{di} and bottom triangle DI$_{id}$ calculated from different alignments (first neighbours and second neighbours pairs) and reconstructing the matrix. Right, comparison of the \textsc{di} and DI$_{id}$ values obtained on the first two pannels.}
 \label{fig:S3}
\end{figurehere}

\section*{Robustness and confidence of the analysis}

\begin{figurehere}
\centering
 \includegraphics{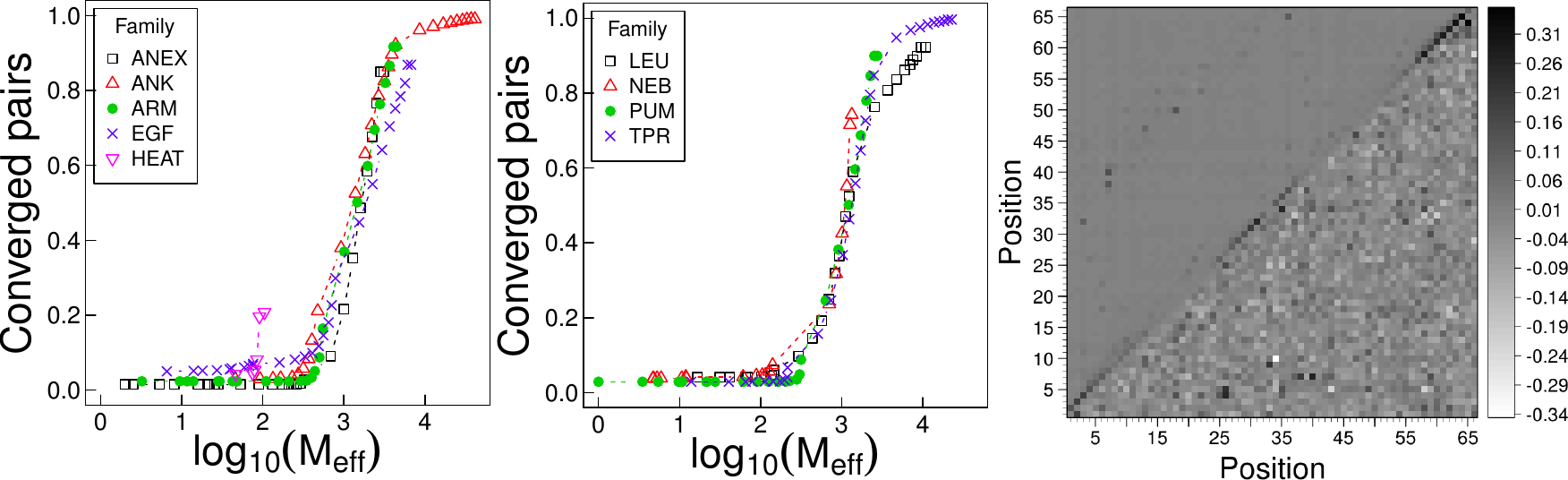}
 \caption{Left and center, for each family proportion of pairs of positions converged accordign to the criteria of the main text vs. the number of effective sequences on the alignment. Right, for the \textsc{ank} family, example of DI$_{id}$ matrix calculated over an alignment of arround 70000 sequences (upper pannel) and over an alignment of arround 400 sequences (lower pannel).}
 \label{fig:S4}
\end{figurehere}

%
\begin{table}
\caption{Repeat protein families analyzed. L is the number of residues of the sequences on the \textsc{fna}; M is the number of sequences on the \textsc{fna}.}
\centering
\begin{tabular}{| lcccc |}
\hline
Family name & Abreviation & Pfam Identifier &  $2 L_0$ & $M$\\
\hline
  \textsc{ankyrin}&\textsc{ank} & PF00023  & 66 & 72908\\
  \textsc{tetratricopeptide}& \textsc{tpr} & PF00515 & 68 &38866\\
  \textsc{leucine rich} &\textsc{leu}& PF00560& 48 &26493\\
  \textsc{spectrin} & \textsc{spec}&PF00435& 212 &13142\\
  \textsc{epidermal growth factor}&\textsc{egf} & PF00008 & 64 &10842\\  
  \textsc{armadillo}& \textsc{arm}& PF00514  &84 &6911\\
  \textsc{annexin} &\textsc{anex}& PF00191& 132 &4264\\
  \textsc{pumilio} &\textsc{pum}& PF00806& 68 &3995\\  
  \textsc{nebulin} &\textsc{neb}& PF00880& 58 &2438\\
  \textsc{heat} & \textsc{heat}&PF02985 & 60 &261\\
  \hline
\end{tabular}
\label{tab:alignments}
\end{table}
\medskip

\end{document}